\documentclass[conference]{IEEEtran}

\usepackage{braket}
\usepackage{amsmath}
\usepackage{graphicx}
\usepackage{mathtools}
\usepackage{url}
\usepackage{bbold}

\newcommand{\subparagraph}{}

\usepackage[compact]{titlesec}
\titlespacing{\section}{0pt}{2ex}{1ex}
\titlespacing{\subsection}{0pt}{1ex}{0ex}
\titlespacing{\subsubsection}{0pt}{0.5ex}{0ex}

\title{Atmospheric Effects on Satellite-to-Ground Quantum Key Distribution using Coherent States}
\author{
	\IEEEauthorblockN{E. Villase\~nor$^1$, R. Malaney$^1$, K. A. Mudge$^2$ and K. J. Grant$^2$}\\
	\IEEEauthorblockA{${}^1$School of Electrical Engineering  \& Telecommunications,\\
		The University of New South Wales, Sydney, NSW 2052, Australia.\\
		${}^2$Defence Science and Technology Group, Edinburgh, SA 5111, Australia.}\\
}

\begin{document}

\maketitle

\IEEEpeerreviewmaketitle

\begin{abstract}
Satellite-based quantum cryptography has already been demonstrated using discrete variable technology. Nonetheless, there is great interest in using weak coherent pulses to perform quantum key distribution (QKD) in the continuous variable (CV) paradigm. In this work, we study the feasibility of performing coherent-state CV-QKD via the satellite-to-ground channel.
We use numerical methods to simulate atmospheric turbulence and compare the results with ground-based experimental data so as to confirm the validity of our approach.
We find the results obtained from the numerical simulations agree well with the experimental data and represent an improvement over the state-of-the-art analytical models. Using the simulation results we then derive QKD key rates and find that useful non-zero key rates can be found over a limited range of zenith angles. Determination of QKD key rates using experimentally validated simulations of low-zenith-angle atmospheric channels represents an important step towards proving the feasibility of real-world satellite-to-Earth CV-QKD.
\end{abstract}

\section{Introduction}
Satellites in space have great potential to accomplish global quantum communications. Compared to fibre-based implementations, the satellite-to-ground free space optical (FSO) channel has considerably lower loss when transmitting quantum signals \cite{spaceQKD}. For space-based quantum communications, a ground-breaking milestone was the proof-of-concept experiments conducted by the satellite {\it Micius} in 2017. Micius consists of a low-Earth-orbit (LEO) satellite equipped with a payload especially designed to perform quantum experiments. Remarkably, Micius proved that the satellite-to-ground FSO channel can be used to distribute quantum entanglement and perform quantum key distribution (QKD) over a distance of 1200km \cite{yin2017satellite, liao2017satellite}.

QKD, arguably the most important quantum communication protocol, provides unconditional security guaranteed by the fundamental laws of quantum mechanics.
In the last decade QKD has advanced from the purely theoretical realm into a well developed technology \cite{NedaReview, QKDReview}. Micius's quantum communications technology is based on the encoding of quantum information in the polarization of single photons, i.e. discrete variable (DV) technology \cite{BB84}.
However, there exists a different paradigm, continuous variable (CV) technology, based on the encoding of quantum information in the quadratures of weak pulses of light \cite{GG02}. CV-QKD shows great promise for increased key rates under the right circumstances, and utilises off-the-shelf well-understood devices, such as homodyne detectors \cite{GaussianQuantumInformation, eps_bound}. Recent experiments lend weight to the viability of satellite-to-ground CV quantum communications \cite{SattelliteAttenuation}, even though  CV-based  quantum keys have only thus far been experimentally distributed  in an FSO channel  over 500m  \cite{.5kmCVQKD}.


There is a need to understand the degradation effects the turbulent atmosphere has on CV quantum signals.
The most relevant analytical model in this regard is the elliptical model of \cite{ellipticalmodel}, which describes the FSO quantum channel under the presence of turbulence (see also \cite{satellite_links, FeasibilityDownlinkCVQKD}).
This model relies on a classical description of the signal to describe the deformations and beam-wandering caused by the
atmosphere. Such an approach has accurately predicted the probabilistic distributions of the transmissivity of the FSO
channel relative to  experiments \cite{ellipticalmodel}.

While the elliptical model describes well the deformations, and wandering of the beam,
the aberrations in the phase wavefront are not fully encompassed in this model. Such aberrations may have an important impact in QKD, as they limit the ability of the signal to interfere with a local oscillator (LO), eventually introducing additional excess noise when doing homodyne or heterodyne measurements \cite{wavefront_aberration}. This is especially important when the LO is generated locally at the receiving station, and a pilot wave is used to synchronise the signal with the LO, the so-called ``local local oscillator'' configuration \cite{FeasibilityDownlinkCVQKD}.
To help alleviate such effects, adaptive optics (AO) techniques have proven useful \cite{AO1, AOQuantum}.

In this work, we use numerical simulations to model the effects of turbulence in the atmosphere to derive the transmissivity of the FSO channel, as well as the wavefront aberrations incurred on the quantum signal.
The contributions of this paper are:
\begin{itemize}
  \item We provide a detailed model of the effects of the atmospheric turbulence on the transmitted quantum signal. Experimental data is used to determine the trustworthiness of this model.
  \item We determine the reduction of excess channel noise by the use of AO to correct wavefront aberrations.
  \item Finally, we provide a realistic determination of the quantum key-rates achievable for the satellite-to-Earth quantum channel.
\end{itemize}

\section{Modelling atmospheric turbulence}
Turbulence in the Earth's atmosphere is caused by random fluctuations in temperature and pressure. These variations alter the air's refractive index both spatially and temporally, distorting any optical waves propagating through the atmosphere. To model the fluctuations the most widely accepted theory was presented by Kolmogorov \cite{Kolmogorov}. In Kolmogorov's theory the turbulence is induced by eddies in the atmosphere characterised by an inner-scale $l_0$, and an outer-scale $L_0$.
The outer-scale denotes the upper-bound to the size of the turbulent eddies. Through dissipative processes larger eddies are transformed into smaller eddies until they reach the size limit of the inner-scale. Below this limit the turbulence is dissipated into the atmosphere as heat.

When using beam propagation to quantify atmospheric turbulence, a useful quantity is the scintillation index ($\sigma_I^2$), defined as the normalised variance of the irradiance fluctuations,
\begin{align}
\sigma_I^2 =\frac{\langle I(x_0, y_0)^2 \rangle }{ \langle I(x_0, y_0) \rangle^2} - 1,
\label{SI}
\end{align}
where $I(x_0, y_0)$ is the optical irradiance evaluated at a single point of the detector plane $\mathcal{D}$, and $\langle \rangle$ the mean over all the measurements performed.
While normally $\sigma_I^2$ is defined using a sole point, here we consider the total power $P$ over  $\mathcal{D}$,
\begin{align}
P = \iint_{\mathcal{D}} |I(x,y)|^2 dxdy,
\end{align}
where $I(x,y)$ is expressed in Cartesian coordinates.
When the detector is small enough, replacing $I(x_0, y_0)$ by $P$ in Eq. \ref{SI} yields a good approximation to $\sigma_I^2$.

To describe the fluctuations of the refractive index we use a spectral density function \cite{book_phase_screen}
\begin{align}
\Phi_{\phi}(\kappa) = 0.49 r_0^{-5/3} \frac{\exp(-\kappa^2/\kappa^2_m)}{{(\kappa^2 + \kappa_0^2)}^{11/6}},
\label{pdf}
\end{align}
with $\kappa$ the radial spatial frequency on a plane orthogonal to the propagation direction, $\kappa_m = 5.92/l_0$, $\kappa_0= 2\pi/L_0$, and $r_0$ the Fried parameter for a propagation length $\Delta z$. In the case of a horizontal propagation path at ground level
\begin{align}
r_0= (0.423 k^2 C_n^2(0) \Delta z)^{-3/5},
\end{align}
with $C_n^2(0)$ the refractive index structure constant at ground level and $k$ the wavenumber.

To model the effects of the atmosphere over a propagating beam we use the phase screen model. The phase screen model consists in subdividing the atmosphere in regions of length $\Delta z$. For each region the random phase changes of the beam are compressed into a thin phase screen, placed at the start of the propagation path, and the rest of the atmosphere is taken to have constant refractive index. In order to simulate beam propagation, we use the software  package {\it PROPER} \cite{proper}, which is an optical propagation library capable of simulating the propagation of electromagnetic waves. The routines in PROPER implement the angular spectrum and Fresnel approximation Fourier algorithms
to propagate a wavefront, in the near-field and far-field conditions, respectively \cite{book_phase_screen}.
In these algorithms the beam is represented using a two-dimensional uniform square grid where each pixel contains a complex number corresponding to the value of the electromagnetic field at that point in space. The phase screens are constructed by performing FFT over a uniform square grid of random complex numbers obtained from a Gaussian distribution with zero mean and variance given by the spectral density function of Eq. \ref{pdf}.

\section{Validating the numerical simulations}
We validate our phase screen-based simulations by first comparing our simulation results to  measured scintillation data.  The atmospheric channel measurements  we adopt in this work were conducted over a 1.5km horizontal path in the DST (Defence Science Technology) Group laser range facility in Adelaide, South Australia. A 7mW laser operating at 1550nm was collimated using a 45mm fibre collimator and measured by three germanium detectors with diameters of 1mm, 5mm and 13mm, respectively. We refer to these tests as ``the experiment'' below.
Additionally, a commercial scintillometer was positioned in parallel to the laser and used to measure $C_n^2(0)$. As presented in \cite{DST_phase_screen},  additional experimental data was used to obtain values for $l_0$ and $L_0$. The experimental values presented in Table \ref{tab:values} serve as inputs in our phase screen simulations.

\begin{table}
\centering
\caption{Experimental parameters}
\begin{tabular}[t]{cccc}
SI & $C_n^2(0)$($\text{m}^{-2/3}$) & $l_0$ (mm) & $L_0$ (m)\\
\hline
3.00 & $2.47\text{e}^{-13}$ & 7.5 & 1.57 \\
\hline
\label{tab:values}
\end{tabular}
\vspace{-10mm}
\end{table}

\begin{figure}
\centering
\includegraphics[width=.40\textwidth]{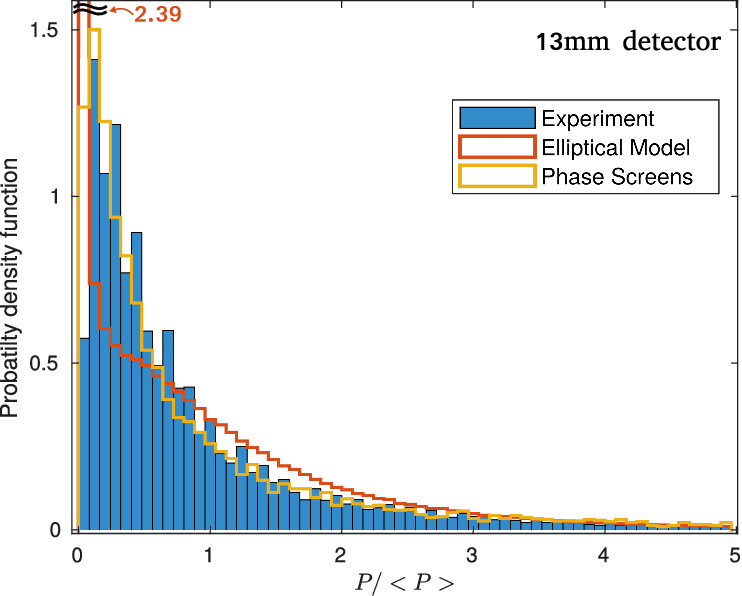}
\caption{Power fluctuations obtained from the experiment, the phase screen simulations and the elliptical model for the 13mm aperture.}
\label{fig:results}
\vspace{-3mm}
\end{figure}

To validate the model we use 10 uniformly spaced phase screens to simulate the turbulence. The grid used in the simulation consists of 1024x1024 points, with a pixel size of 1.1mm. The beam waist and propagation length of the beam, and positions of the detectors are consistent with those adopted in the experiment.
The results, presented in Fig. \ref{fig:results}, show that the probability density function (PDF) of $P/\langle P \rangle$ obtained from our simulations matches to a good degree the distribution obtained from the experiment.  Moreover, we see that the simulation is a slightly better match to the experimental data than that obtained via the analytical elliptical model. For brevity we present only the results for the 13mm diameter aperture, but we note a similar behaviour is observed for the 1mm and 5mm apertures. Additionally, we present in Table \ref{tab:si} the values of $\sigma_I^2$ obtained for every detector size. The value $\sigma_I^2$ determined was obtained from a sample of 10000 simulations, with an accuracy greater than $90\%$, as discussed in \cite{DST_phase_screen}. We see that the simulation values agree within statistical error with the measurement results. On the other hand, we see that the values obtained from the elliptical model are considerably different. This is as expected, since the elliptical model does not account for the small variations in intensity related to scintillation \cite{ellipticalmodel}.

\begin{table}
\centering
\caption{Scintillation index ($\sigma_I^2$)}
\begin{tabular}[t]{c|ccc}
Detector & Experiment & Simulation & Elliptical \\
\hline
1mm & 3.02 & 3.31 & 1.52  \\
5mm & 3.00 & 3.13 & 1.52  \\
13mm & 2.22 & 2.32 & 1.50
\label{tab:si}
\end{tabular}
\vspace{-5mm}
\end{table}

\section{Modelling the Earth-satellite channel}
With our ground-based simulations validated, our main objective now is to model the atmospheric turbulence of the satellite-to-ground channel. We consider a LEO satellite, corresponding to an altitude between 300km to 1000km. To obtain the refractive index structure of the atmosphere we use the widely used Hufnagel-Valley model \cite{HVmodel}:
\begin{align}
C_n^2(h) &= 0.00594(v/27)^2 (10^{-5}h)^{10} \exp(-h/1000) \\ \nonumber
          &+ 2.7\times10^{-16} \exp(-h/1500) + A \exp(-h/100),
\end{align}
with $h$ the altitude in meters, $v=21$ the rms wind-speed ($m/s$), and $A=1.7 \times 10^{-14}$ the nominal value of $C_n^2(0)$ at the ground.
Additionally, measurements made of the scintillation suggest the outer scale $L_0$ changes with the altitude according to the empirical Coulman-Vernin profile \cite{outer_scale}
\begin{align}
L_0(h) = \frac{4}{1 + (\frac{h-8500}{2500})^2},
\end{align}  a function we adopt.
We also set the inner-scale to be a fraction of the outer-scale, as $l_0=0.005L_0$.

\begin{figure}
\centering
\includegraphics[width=.47\textwidth]{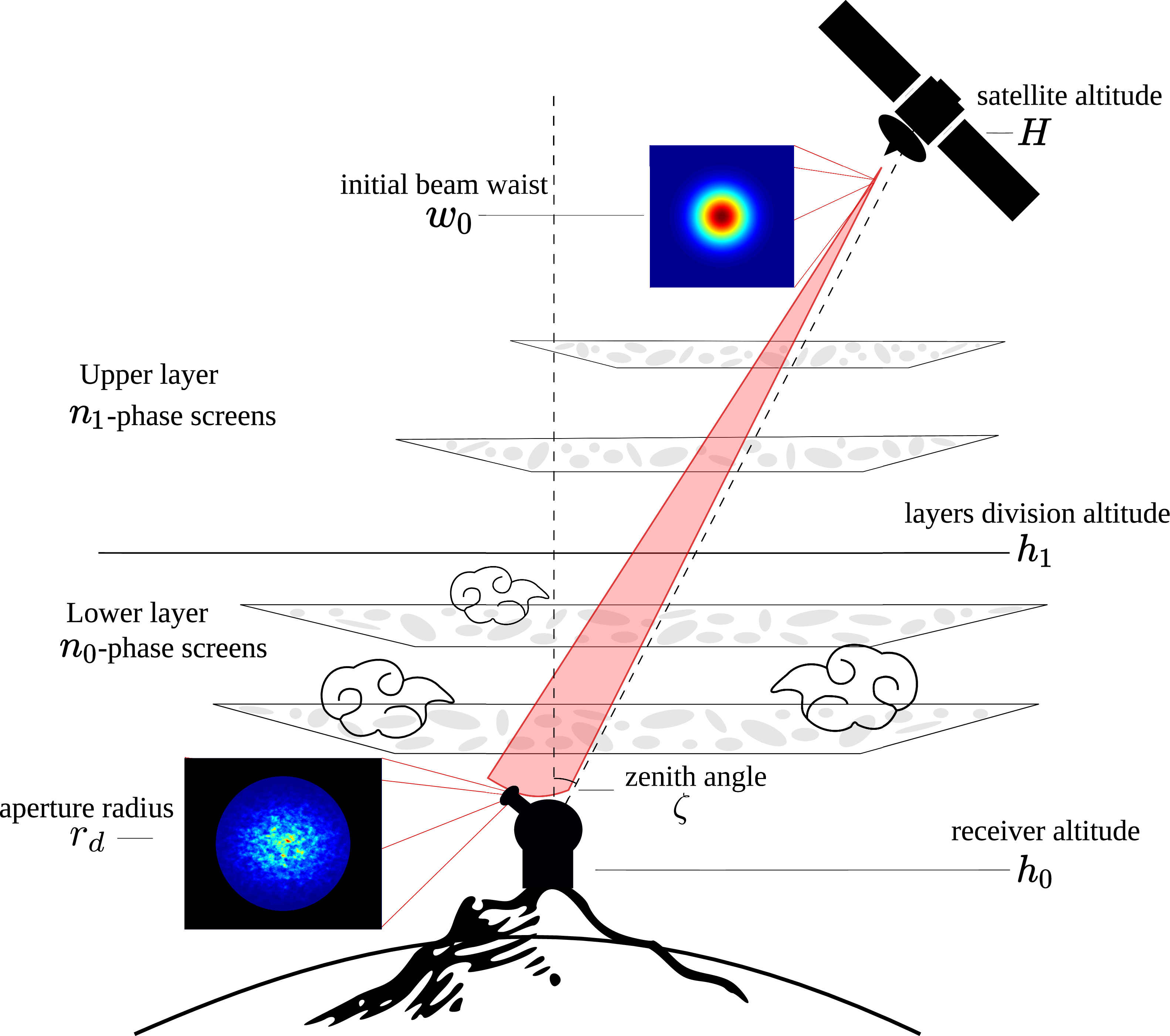}
\caption{Phase screens system to model a satellite-to-ground channel. The atmosphere is divided in two layers, with the lower layer containing most of the turbulence effects.}
\label{fig:downlink}
\vspace{-3mm}
\end{figure}

With all the above considerations, we use the system presented in Fig. \ref{fig:downlink} to simulate the atmospheric effects of a satellite-to-ground channel from a satellite at an altitude $h=H$ to a ground station at an altitude $h_0$. In this system the atmosphere is divided in two layers at an altitude $h_1$. A number $n_1$ and $n_0$ of phase screens are positioned at equal intervals for each one of the upper and lower layers, respectively. Most of the turbulence is contained in the lower layer, therefore, $n_0 > n_1$. The quantum signal initially possesses a diffraction-limited Gaussian intensity profile with beam waist $w_0$. The signal is detected by the ground station with an aperture of radius $r_d$.  The total path the signal has to travel from the satellite to the ground station depends on the zenith angle $\zeta$. We have not considered the elongation of the path due to the refractive effects of the atmosphere, but we expect this factor to be approximately 1.05 for $\zeta=75^{\circ}$ \cite{satellite_links}. Besides the signal, we consider a strong local oscillator (LO) is generated at the ground stations for the purpose of performing homodyne measurements.
The parameters used in the simulation system are presented in Table \ref{tab:params}.
For each $\zeta$ we corroborate that the positions of the phase screens satisfy the condition that $<10\%$ of the
total scintillation is allowed to take place over the
distance between phase screens \cite{downlink_simulations_10perc}.
\begin{table}
\centering
\caption{Satellite-to-ground simulation parameters}
\begin{tabular}[t]{ccccccc}
$H$   &
$h_0$ &
$h_1$ &
$n_0$ &
$n_1$ &
$w_0$ &
$\lambda$ \\
\hline
300km &
2km &
20km &
10 &
1 &
15cm &
1550nm \\
\hline
\label{tab:params}
\end{tabular}
\vspace{-5mm}
\end{table}

The scintillation for a satellite-to-ground channel with weak turbulence can be calculated using \cite{andrews_book1}
\begin{align}
  \sigma_I^2 = 2.25 k^{7/6} \sec^{11/6} (\zeta) \int_{h_0}^H C_n^2(h) (h -h_0)^{5/6} dh,
  \label{eq:scint_downlink}
\end{align}
We test our system by comparing the scintillation index obtained from the simulation with the analytical expression Eq. \ref{eq:scint_downlink}.
As before, we use a grid that consists of 1024x1024 points, now with a pixel size of 7.8mm. For each set of parameters, we execute the simulations 10000 times.
The Fried parameter used in Eq. \ref{pdf} is now \cite{andrews_book1}
\begin{align}
r_0 = \Big( 0.423 k^2 \sec(\zeta)\int_{h^-}^{h^+} C_n^2(h) dh \Big)^{-3/5},
\end{align}
where $h^-$ and $h^+$ correspond to the lower and upper altitudes of the propagation path corresponding to the respective phase screen.
In the simulation, the scintillation is measured from the pixel at the centroid of the receiver plane.
The resulting values, shown in Table \ref{tab:si_downlink}, agree with each other within some margin of error, therefore corroborating the simulation system. For the remaining results presented in this work, the parameters and the grid characteristics will remain unchanged.
\begin{table}
\centering
\caption{Scintillation index ($\sigma_I^2$) - Satellite-to-ground}
\begin{tabular}[t]{c|cc}
$\zeta$ (deg) & Theory & Simulation  \\
\hline
$0^{\circ}$ & 0.033 & 0.034  \\
$10^{\circ}$ & 0.034 & 0.035  \\
$20^{\circ}$ & 0.038 & 0.039  \\
$30^{\circ}$ & 0.043 & 0.045  \\
$40^{\circ}$ & 0.053 & 0.056  \\
$50^{\circ}$ & 0.071 & 0.078  \\
$60^{\circ}$ & 0.113 & ~0.123
\label{tab:si_downlink}
\end{tabular}
\vspace{-5mm}
\end{table}

\subsection{Adaptive optics to correct wavefront}
To generate a correction to the quantum signal in the simulations,
 we assume a beacon beam is available to characterise the turbulence effects and provide feedback to the AO element used to correct the wavefront aberrations in the signal.
The beacon beam is such that before being disturbed by the atmosphere it corresponds to a plane wave with a constant intensity profile, such as light from a distant star.
The characterisation is made by projecting the beacon wavefront aberrations into a basis of orthogonal polynomials in a plane disk, known as Zernike polynomials \cite{AO1}. The projection in the Zernike basis is then used to construct a correction, which is applied by means of a deformable mirror \cite{AO1}.
We briefly describe this process in more detail as follows.

We can quantify the aberrations of the wavefront caused by the turbulence using the {\it coherent efficiency}, defined as \cite{AOQuantum}
\begin{align}
  \gamma = \frac{|\frac{1}{2} \iint_{\mathcal{D}}[ E_\text{ref}^* E_\text{beacon} +  E_\text{ref} E_\text{beacon}^*  ] ds|^2}{\iint_{\mathcal{D}} |E_\text{ref}| ^2 ds \iint_{\mathcal{D}}|E_\text{beacon}|^2 ds},
\end{align}
with $E_\text{beacon}$ is the electric field of the beacon, and $E_\text{ref}$ is a reference wave that remains undisturbed by the turbulence. A value of $\eta=1$ corresponds to a perfect alignment between $E_\text{beacon}$ and $E_\text{ref}$.
The wavefront aberrations of the signal will introduce additional excess noise to the quantum signal as \cite{AOQuantum}
\begin{align}
  \xi_\text{det}(\gamma) =  \frac{((1 - \gamma) + \upsilon_\text{el})\eta_\text{det}}{\gamma},
  \label{AOnoise}
\end{align}
where $\upsilon_\text{el}$ is the electronic noise inherent to the measurement devices (including the AO system) and $\eta_\text{det}$ the detector efficiency.
As discussed in the next section the value of $\xi_\text{det}$ has an impact on the the effectiveness of CV-QKD.

The Zernike polynomials are defined, in polar coordinates $r$ and $\phi$, as
\begin{align}
   Z_n^m(r, \phi)=
    \begin{cases}
      R_n^m(r) \cos(m \phi), & \text{if}\ m \geq 0 \\
       R_n^{-m}(r) \sin(-m \phi), & \text{otherwise},
    \end{cases}
\end{align}
where $m$ and $n$ are integers and
\begin{align}
	R_n^m(r) = \sum_{k=0}^{\frac{n-m}{2}} \frac{(-1)^k (n-k)!}{k! (\frac{n+m}{2} -k)! (\frac{n-m}{2} -k)!} r^{n-2k},
\end{align}
for $n - m$ even, and $R_n^m = 0$ for $n - m$ odd.
Using the phase wavefront of the beacon beam $\Phi=\text{arg}(E_{\text{beacon}})$, where $\text{arg}$ is the complex argument function, a correction $C$ can be constructed as
\begin{align}
C(r, \phi) &= \sum_{n}^{n_\text{max}} \sum_{m=-n}^{n} a_{m,n} Z_n^m(r, \phi),\\ \nonumber
a_{m,n} &= \frac{2n+2}{\epsilon_m \pi} \int_0^\infty \int_0^{2\pi} \Phi(r, \phi) Z_n^m(r,\phi) rdr d\phi,
\end{align}
where $n_\text{max}$ is the maximum order of the polynomials used in the correction, and $\epsilon_m=2$ if $m=0$ and $\epsilon_m=1$ otherwise.

\begin{figure}
\centering
\includegraphics[width=.45\textwidth]{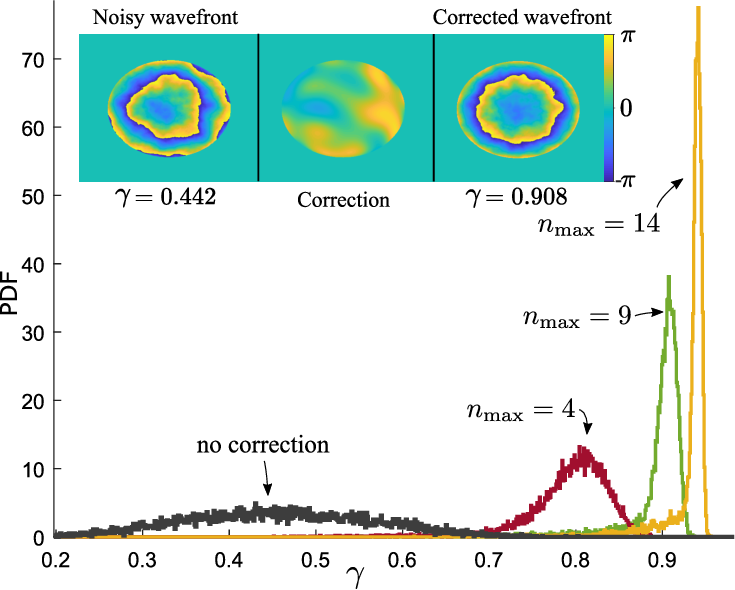}
\caption{PDF of the coherence efficiencies for the satellite-to-ground channel with $\zeta=0$ and $r_d=1$. For the same input signal, AO corrections with different maximum ranges of Zernike polynomials are applied. (inset) Example of the phase wavefront of a signal before and after correction.}
\label{fig:phase}
\vspace{-3mm}
\end{figure}

Under the assumption that the AO is capable of adjusting at the required frequency to compensate for the fluctuations in time of the atmosphere.
The effectiveness of the correction ultimately depends on the value $n_\text{max}$. Ideally, we desire $n_\text{max}$ to be as large as possible so as to be able to address the smallest aberrations of the wavefront. To analyse the effectiveness of the corrections for the satellite-to-ground channel, in Fig. \ref{fig:phase} we show the PDFs of $\gamma$ obtained for different values of $n_\text{max}$. We see that higher values of $n_\text{max}$ greatly increase the values of $\langle \gamma \rangle$.

\subsection{Transmissivity of the satellite-to-ground channel}
Using the simulations we can find the PDF of the transmissivity of the satellite-to-ground FSO channel. The transmissivity is calculated from the total power at the receiver. We use the normalised (unitless) power $P'$ of the signal, corresponding to the power at the receiver divided by the transmitted optical power at the exit of
the transmitter, $P'= P/P_0$ (we assume the light source is constant).
To account for pointing errors between the satellite and ground station, and the absorption of the optical signal by the atmosphere, we consider a fixed loss $T'=2$dB \cite{liao2017satellite}. Additionally, we consider a detector efficiency of $\eta_\text{det} =1$dB. Therefore we model the entire transmissivity of the satellite-to-ground channel as $T = P' \times T' \times \eta_\text{det}$.

\begin{figure}
\centering
\includegraphics[width=.45\textwidth]{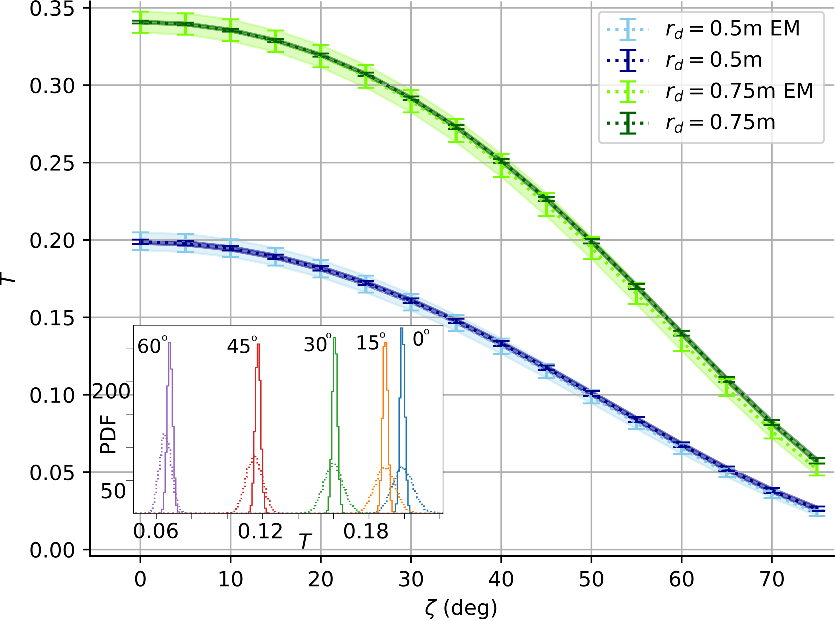}
\caption{Mean values of the transmissivities obtained from the phase screen simulations, and from the elliptical model (EM) for two aperture radius, the thickness of the lines correspond to the standard deviations. (inset) PDFs of the simulations (solid line), and the elliptical model (dashed line) for selected zenith angles, and $r_d=0.5$m.}
\label{fig:res_downlink}
\end{figure}

Using the simulations we obtain the PDF of $T$ for different values of $\zeta$ and two different receiver aperture sizes. We compare the values obtained with the ones obtained from the elliptical model under the same conditions, the results are summarised in Fig. \ref{fig:res_downlink}.
We observe that compared to our simulations the PDFs obtained from the elliptical model are slightly pessimistic, since they present greater variances for each zenith angle.

\section{Satellite-based CV-QKD}
 We consider QKD is achieved between the satellite and the ground station using the protocol {\it GG02}, introduced by Grosshans and Grangier in 2002 \cite{GG02}.
 In this protocol the sender $A$ prepares Gaussian modulated coherent states which are measured by the receiver $B$ using homodyne detection.
The key rate depends on three parameters, the modulated variance of the coherent states $V_{\text{mod}}$, the total transmissivity of the channel $T$, and the excess quantum noise incurred during the protocol $\xi$ (expressed in vacuum noise units). The excess noise is obtained as $\xi= \xi_\text{ch} + \xi_\text{det}/T$, with the individual noise components from the atmospheric channel $\xi_\text{ch}$, and the noise of the measurement devices $\xi_\text{det}$, as defined in Eq. \ref{AOnoise}.
Due to the fluctuating nature of the satellite-to-ground channel the parameters $T$ and $\xi$ are described by PDFs. Therefore, as discussed in \cite{NedaComposable}, we need to consider the ensemble-averages when doing the security analysis to calculate the key rates. Alternatively, the analysis can be derived as in the non-fluctuating channel if we define an {\it effective transmissivity} $T_{f}$, and an {\it effective excess noise} $\xi_f$, as
\begin{align}
  &T_{f} = \langle \sqrt{T} \rangle^2 ~~~~~~~
  T_{f} \xi_f = \text{Var}(\sqrt{T}) V_\text{mod} + \langle T \xi \rangle \\ \nonumber
  &\text{Var}(\sqrt{T}) = \langle T \rangle - \langle \sqrt{T} \rangle^2,
\end{align}
with the mean values computed as
\begin{align}
  &\langle T \rangle = \int_0^1 T p_\zeta(T) dT  ~~~~~
  \langle \sqrt{T} \rangle = \int_0^1 \sqrt{T} p_\zeta(T) dT \\ \nonumber
  &\langle T \xi \rangle = \xi_\text{ch}   \langle T \rangle + \int_0^1 \xi_\text{det}(\gamma) p_\zeta (\gamma) d\gamma,
\end{align}
with $p_\zeta(T)$ and $p_\zeta(\gamma)$ the PDFs of $T$ and $\gamma$ for a given $\zeta$, respectively.

Following the procedure in \cite{PracticalCVQKD}, the key rate under reverse reconciliation is computed as
\begin{align}
 K = \beta I_{AB} - \chi_{BE},
\end{align}
where $\beta$ is the reverse reconciliation efficiency, $I_{AB}$ the shared information between satellite and ground station, and $\chi_{BE}$ the Holevo information acquired by the eavesdropper.
The value of $I_{AB}$ is directly related to the signal-to-noise ratio (SNR) of the quantum signal. For the GG02 protocol we have
\begin{align}
I_{AB} = \frac{1}{2} \log_2(1 + \text{SNR}) = \frac{1}{2} \log_2 \Big(1 + \frac{T_{f} V_{\text{mod}}}{1+T_{f}\xi_f}\Big).
\end{align}

For simplicity in the calculation of the Holevo information, we perform the security analysis as in the entanglement based (EB) version of the GG02 protocol. This security analysis applies to GGO2 since both protocols are equivalent \cite{PracticalCVQKD}.
In the EB version of the protocol the covariance matrix of the state after it has been received by the ground station, expressed in terms of $V_{\text{mod}}$, is
\begin{align}
M_{AB} &= \begin{pmatrix}
a \mathbb{1} & c\sigma_z \\
c\sigma_z & b \mathbb{1}
\end{pmatrix} \\
&=  \begin{pmatrix}
(V_{\text{mod}}+1) \mathbb{1} & \sqrt{T_{f} (V_{\text{mod}}^2 + 2V_{\text{mod}})}\sigma_z \\
\sqrt{T_f (V_{\text{mod}}^2 + 2V_{\text{mod}})}\sigma_z & (T_{f}V_{\text{mod}} + 1 +T_{f}\xi_f) \mathbb{1} \nonumber
\end{pmatrix},
\end{align}
where $\mathbb{1}=\text{diag}(1,1)$, and $\sigma_z=\text{diag}(1,-1)$.
We presume the eavesdropper holds a purification of the shared quantum state. This means that the Holevo information is
\begin{align}
\chi_{BE} = S_{AB} - S_{A|B},
\end{align}
where $S$ is the von Neumman entropy
\begin{align}
S(\rho) = \sum_i g(\nu_i).
\end{align}
For a given state $\rho$, the value of $S$ is calculated from the symplectic eigenvalues $\{\nu_i \}$ of the covariance matrix of $\rho$, and the function $g(x)$ is
\begin{align}
g(x) = \frac{x+1}{2} \log_2\Big(\frac{x+1}{2}\Big) -  \frac{x-1}{2} \log_2\Big(\frac{x-1}{2}\Big).
\end{align}
For the covariance matrix $M_{AB}$ it is straightforward to show its eigenvalues are
\begin{align}
\nu_{1,2} = \frac{1}{2} (z \pm [b-a]) , ~~~~~
z = \sqrt{(a+b)^2 - 4c^2}.
\end{align}
When homodyne measurement is used the symplectic eigenvalue of $M_{A|B}$ is
\begin{align}
\nu_3 = \sqrt{a\Big(a -\frac{c^2}{b}\Big)}.
\end{align}
We only consider homodyne measurements, since it has been shown that when the value of $T_f$ is low, the key rates achieved using homodyne measurements are higher compared to the ones obtained using heterodyne measurements \cite{AtmosphericQKD}.

%

\begin{figure}
\centering
\includegraphics[width=.45\textwidth]{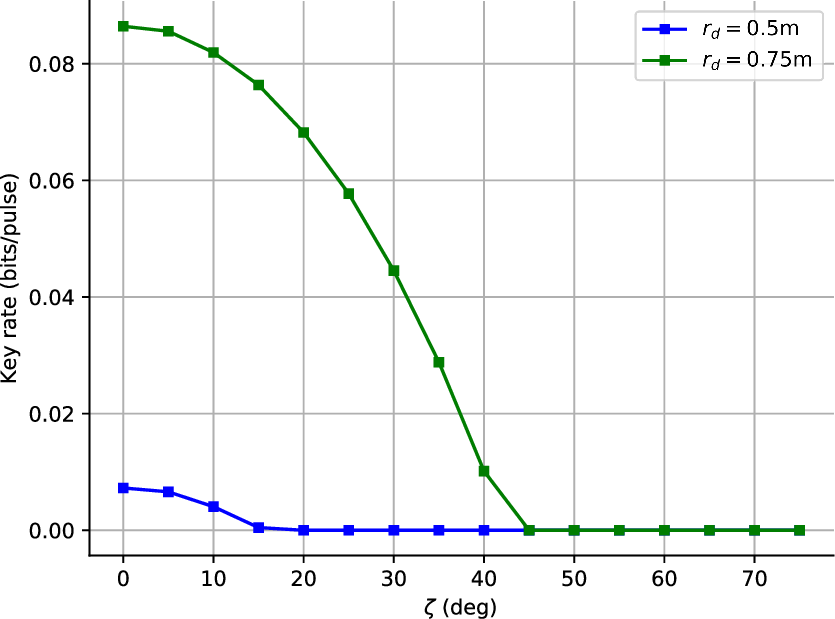}
\caption{Key rates for CV-QKD via the satellite-to-ground channel as a function of the zenith angle, $\zeta$, for different values of $r_d$.}
\label{fig:kr}
\vspace{-3mm}
\end{figure}

Using the PDFs of both $T$ and $\xi_\text{det}$, obtained from the simulations for different values of $\zeta$, we compute the key rates for the satellite-to-ground channel. We include AO corrections with the values of $n_\text{max}=9$ for $r_d=0.5$, and $n_\text{max}=14$ for $r_d=0.75$, and use the obtained values of $\gamma$ to calculate the excess noise added by wavefront aberrations. The remaining excess noise parameters are set to $\xi_\text{ch}= 0.02$, and $\upsilon_\text{el}= 0.005$. Additionally, we set $\beta=0.95$, a value that can be achieved using modern techniques \cite{revere_reconciliation}.
The results, shown in Fig. \ref{fig:kr}, indicate that high key rates can be achieved for the satellite-to-ground channel for low zenith angles. While for  an aperture radius of $r_d=0.5$ non-zero key rates are limited to the lowest zenith angles, for an aperture radius $r_d=0.75$ non-zero key rates can be obtained for zenith angles up to $40^{\circ}$.

\section{Conclusions}
In this work we analysed the key rates obtained using the satellite-to-ground channel for CV-QKD based on Gaussian modulated coherent states.
Our analysis is based on numerical simulations of beam propagation from a satellite to a ground station through a turbulent atmosphere. We validated our simulations using measurements from a laser propagation experiment over a horizontal channel.

The resulting key rates show that the effects of turbulence in the atmosphere are not an impediment in reaching high key rates for  low values of zenith angles. The caveat, however, is that advanced satellite technology is required, most notably large detector apertures with integrated adaptive optics, and advanced satellite pointing systems. Our results motivate further experiments of satellite-based CV-QKD.

This research was a collaboration between the Commonwealth of Australia (represented by the Defence Science and Technology Group) and the University of New South Wales through a Defence Science Partnerships agreement.

\bibliographystyle{unsrt2}

\begingroup
\raggedright

\bibliography{bib}

\begin{thebibliography}{10}

\bibitem{spaceQKD}
J.~G. Rarity et~al.
\newblock Ground to satellite secure key exchange using quantum cryptography.
\newblock {\em New Journal of Physics}, 4 : 82--82, 2002.

\bibitem{yin2017satellite}
J.~Yin et~al.
\newblock Satellite-based entanglement distribution over 1200 kilometers.
\newblock {\em Science}, 356 (6343)  : 1140--1144, 2017.

\bibitem{liao2017satellite}
S.~Liao et~al.
\newblock Satellite-to-ground {QKD}.
\newblock {\em Nature}, 549 : 43--47, 2017.

\bibitem{NedaReview}
N.~{Hosseinidehaj} et~al.
\newblock Satellite-{B}ased {CV} quantum communications: State-of-the-art and a
  predictive outlook.
\newblock {\em IEEE Communications Surveys Tutorials}, 21 (1)  : 881--919,
  2019.

\bibitem{QKDReview}
S.~Pirandola et~al.
\newblock Advances in quantum cryptography.
\newblock {\em Advances in Quantum Cryptography}, DOI:10.1364/AOP.361502, 2019.

\bibitem{BB84}
C.~H. Bennett and G.~Brassard.
\newblock {Quantum cryptography: Public key distribution and coin tossing}.
\newblock {\em Theoretical Computer Science}, 560 : 7--11, 2014.

\bibitem{GG02}
F.~Grosshans and P.~Grangier.
\newblock {CV} quantum cryptography using coherent states.
\newblock {\em Phys. Rev. Lett.}, 88 (5)  : 057902 (4), 2002.

\bibitem{GaussianQuantumInformation}
C.~Weedbrook et~al.
\newblock {G}aussian quantum information.
\newblock {\em Rev. Mod. Phys.}, 84 (2)  : 621--669, 2012.

\bibitem{eps_bound}
V.~{Usenko} et~al.
\newblock Continuous- and discrete-variable {QKD} with nonclassical light over
  noisy channels.
\newblock In {\em International Conference on Telecommunications and Signal
  Processing},  753--756, 2016.

\bibitem{SattelliteAttenuation}
K.~G\"{u}nthner et~al.
\newblock Quantum-limited measurements of optical signals from a geostationary
  satellite.
\newblock {\em Optica}, 4 (6)  : 611--616, 2017.

\bibitem{.5kmCVQKD}
S.~Y. Shen et~al.
\newblock Free-space {CV-QKD} of unidimensional {G}aussian modulation using
  polarized coherent states in an urban environment.
\newblock {\em Phys. Rev. A}, 100 (1)  : 012325 (8), 2019.

\bibitem{ellipticalmodel}
D.~Vasylyev et~al.
\newblock Atmospheric quantum channels with weak and strong turbulence.
\newblock {\em Phys. Rev. Lett.}, 117 (9)  : 090501 (6), 2016.

\bibitem{satellite_links}
D.~Vasylyev et~al.
\newblock Satellite-mediated quantum atmospheric links.
\newblock {\em Phys. Rev. A}, 99 (5)  : 053830 (27), 2019.

\bibitem{FeasibilityDownlinkCVQKD}
D.~Dequal et~al.
\newblock Feasibility of satellite-to-ground {CV-QKD}.
\newblock {\em arXiv:2002.02002}, 2020.

\bibitem{wavefront_aberration}
K.~A. Winick.
\newblock Atmospheric turbulence-induced signal fades on optical heterodyne
  communication links.
\newblock {\em Appl. Opt.}, 25 (11)  : 1817--1825, 1986.

\bibitem{AO1}
L.~Zhu et~al.
\newblock Wave-front generation of {Z}ernike polynomial modes with a
  micromachined membrane deformable mirror.
\newblock {\em Appl. Opt.}, 38 (28)  : 6019--6026, 1999.

\bibitem{AOQuantum}
Y.~Wang et~al.
\newblock Performance improvement of free-space {CV-QKD} with an adaptive
  optics unit.
\newblock {\em Quantum Information Processing}, 18 (251)  : 251 (21), 2019.

\bibitem{Kolmogorov}
A.~N. Kolmogorov et~al.
\newblock The local structure of turbulence in incompressible viscous fluid for
  very large {R}eynolds numbers.
\newblock {\em Proceedings of the Royal Society of London}, 434 (1890) , 1991.

\bibitem{book_phase_screen}
J.~D. Schmidt.
\newblock {\em Numerical Simulation of Optical Wave Propagation with Examples
  in MATLAB}.
\newblock SPIE press, 2010.

\bibitem{proper}
J.~E. Krist.
\newblock {PROPER: an optical propagation library for IDL}.
\newblock In {\em Optical Modeling and Performance Predictions III}, volume
  6675,  250--258. SPIE, 2007.

\bibitem{DST_phase_screen}
K.~A. {Mudge} et~al.
\newblock Scintillation index of the free space optical channel: Phase screen
  modelling and experimental results.
\newblock In {\em International Conference on Space Optical Systems and
  Applications},  403--409, 2011.

\bibitem{HVmodel}
R.~E. Hufnagel and N.~R. Stanley.
\newblock Modulation transfer function associated with image transmission
  through turbulent media.
\newblock {\em J. Opt. Soc. Am.}, 54 (1)  : 52--61, 1964.

\bibitem{outer_scale}
C.~E. Coulman et~al.
\newblock Outer scale of turbulence appropriate to modeling refractive-index
  structure profiles.
\newblock {\em Appl. Opt.}, 27 (1)  : 155--160, 1988.

\bibitem{downlink_simulations_10perc}
J.~M. Martin and S.~M. Flatt\'{e}.
\newblock Intensity images and statistics from numerical simulation of wave
  propagation in 3-d random media.
\newblock {\em Appl. Opt.}, 27 (11)  : 2111--2126, 1988.

\bibitem{andrews_book1}
L.~C. Andrews and R.~L. Phillips.
\newblock {\em Laser Beam Propagation through Random Media}.
\newblock SPIE press, second edition, 2005.

\bibitem{NedaComposable}
N.~Hosseinidehaj et~al.
\newblock Composable finite-size effects in free-space {CV-QKD} systems.
\newblock {\em arXiv:2002.03476}, 2020.

\bibitem{PracticalCVQKD}
F.~Laudenbach et~al.
\newblock {CV-QKD} with {G}aussian modulation - the theory of practical
  implementations.
\newblock {\em Advanced Quantum Technologies}, 1 (1)  : 1800011 (37), 2018.

\bibitem{AtmosphericQKD}
S.~Wang et~al.
\newblock Atmospheric effects on {CV-QKD}.
\newblock {\em New Journal of Physics}, 20 (8)  : 083037 (20), 2018.

\bibitem{revere_reconciliation}
F.~Furrer.
\newblock Reverse-reconciliation {CV-QKD} based on the uncertainty principle.
\newblock {\em Phys. Rev. A}, 90 (4)  : 042325 (12), 2014.

\end{thebibliography}
\endgroup

\end{document}